\documentclass[12pt]{article}

\usepackage{amsmath,amsthm,amssymb,makeidx,mathrsfs,chemarr}         %
\usepackage{graphicx}        %
\usepackage[margin=1in]{geometry}
\usepackage{multicol}        %
\usepackage[bottom]{footmisc}%

\newtheorem{theorem}{Theorem}
\newtheorem{proposition}{Proposition}
\newtheorem{definition}{Definition}
\newtheorem{corollary}{Corollary}
\newtheorem{remark}{Remark}
\makeindex             %

\begin{document}
	
	\title{\Large \bfseries Analysis of a reduced model of epithelial–mesenchymal fate determination in cancer metastasis as a singularly-perturbed monotone system}
	
	\author{\large M. Ali Al-Radhawi$^1$ and Eduardo D. Sontag$^{1,2}$}
	\date{\small $^{\rm 1}$Departments of Bioengineering and Electrical and Computer Engineering, , Northeastern University, 360 Huntington Avenue, Boston MA 02115. \\ $^{\rm 2}$Laboratory of Systems Pharmacology, Program in Therapeutic Science, Harvard Medical School, Boston MA 02115. \\ Emails: \texttt{\{malirdwi,e.sontag\}@northeastern.edu} \\ \strut \\ \today}

\maketitle

\begin{abstract}Tumor metastasis is one of the main factors responsible for the high fatality rate of cancer. Metastasis can occur after malignant cells transition from the epithelial phenotype to the mesenchymal phenotype. This transformation allows cells to migrate via the circulatory system and subsequently settle in distant organs after undergoing the reverse transition  {from the mesenchymal to the epithelial phenotypes}. The core gene regulatory network controlling these transitions consists of a system made up of coupled SNAIL/miRNA-34 and ZEB1/miRNA-200 subsystems.   In this work, we formulate a mathematical model   {of the core regulatory motif} and analyze its long-term behavior. We start by developing a detailed reaction network with 24 state variables. Assuming fast promoter and mRNA kinetics, we then show how to reduce our model to a monotone four-dimensional system.  For the reduced system, monotone dynamical systems theory can be used to prove generic convergence to the set of equilibria for all bounded trajectories.  The theory does not apply to the full model, which is not monotone, but we briefly discuss results for singularly-perturbed monotone systems that provide a tool to extend convergence results from reduced to full systems, under appropriate time separation assumptions.\end{abstract}

 \section{Introduction}
Realistic dynamical models of physical systems are often very complex and high-dimensional, and this is especially so in molecular cell biology.  Effective analysis often requires simplifications through model reduction techniques.  A traditional approach to model reduction in biology is to take advantage of time-scale separation.  Indeed, most interesting processes in biology are made up of subsystems which operate at different time scales, thus allowing fast subprocesses to be ``averaged out'' at the observational time scale. The rigorous mathematical analysis of time scale separation,
and in particular the mathematical field of \emph{singular perturbations}, owe much to the pioneering work of Tikhonov~\cite{tikhonov52}.  Singular perturbation theory now plays a key role in both science and engineering~\cite{kokotovic99}.

There are many examples of the ubiquitousness of time-scale separation in molecular biology.  An early example is the study of enzymatic reactions, where the derivation of \emph{Michaelis-Menten} kinetics in 1913 \cite{michaelis1913} is still widely used \cite{gunawardena14}.  Another example is provided by Gene Regulation Networks (GRNs) which naturally have multiple levels of time scales: external stimuli change Transcription Factor (TF) activities in milliseconds, promoter kinetics equilibrate in seconds, transcription and translation take minutes, and protein kinetics are in the order of tens of minutes to hours~\cite{alon06}.  

\begin{figure}[ht]
	\centering
	\includegraphics[width=0.5\textwidth]{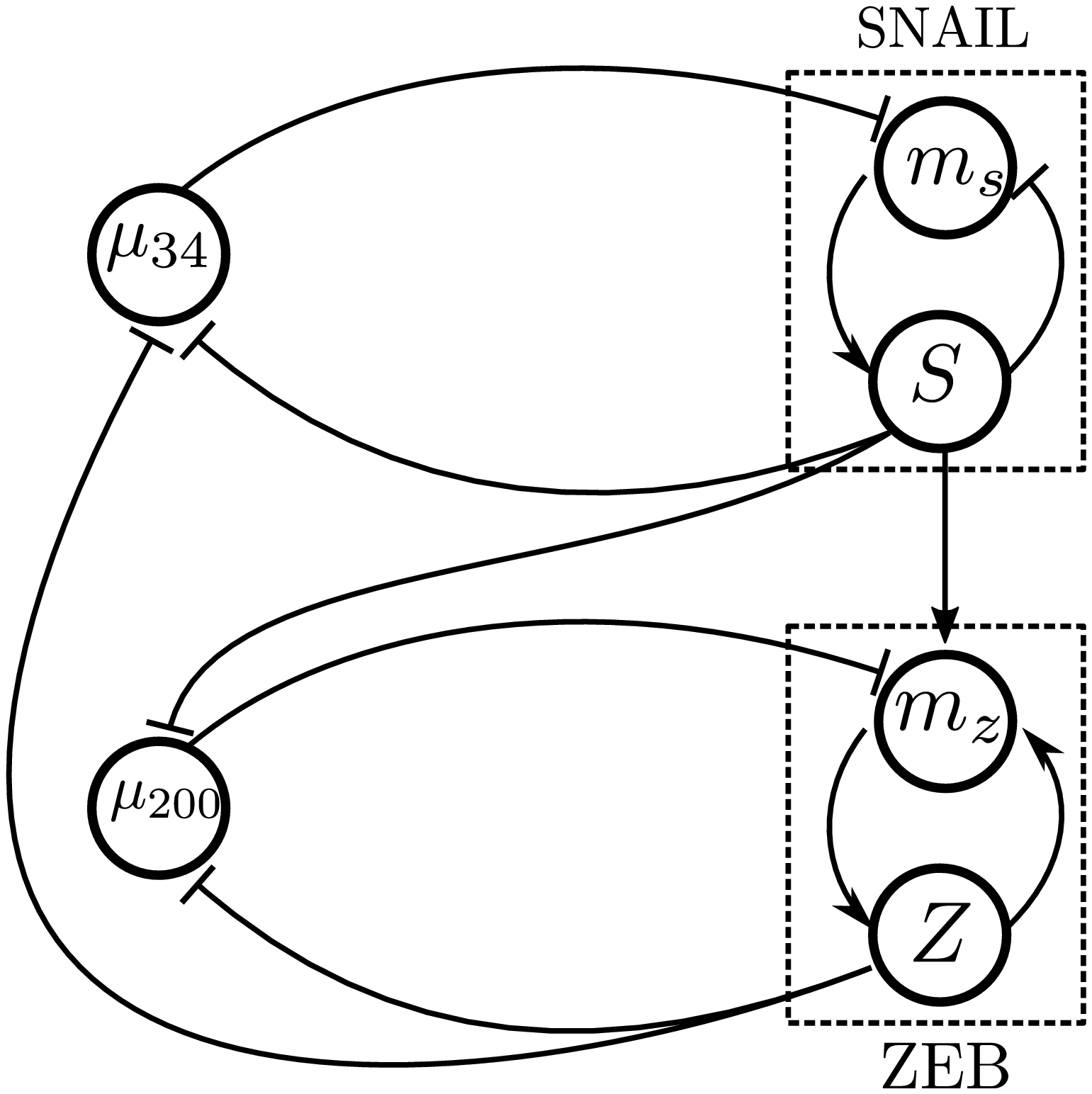}
	\caption{The core EMT/MET network \cite{levine13}, which involves  {proteins, mRNAs, microRNAs (miRNAs) and the underlying genes which are not explicitly depicted}.   The mRNAs of SNAIL and ZEB1 are denoted by $m_s, m_z$, respectively. An arrow of the form ``$\rightarrow$'' denotes activation, while ``$\dashv$'' denotes inhibition.   The detailed reaction network model is presented in \S 4. }
	\label{f.circuit}
\end{figure}

In this paper we study a GRN that determines cell-fate in the metastasis of cancerous tumors. This network regulates a transition between two cell types: \emph{epithelial cells}, which line the external and internal surfaces of many organs, and \emph{mesenchymal stem cells}, which are multipotent connective tissue cells that can differentiate into other type of cells such as muscles, bone, etc. Bidirectional transitions between these two cell types can happen, and they are referred to as ``Epithelial to Mesenchymal'' transitions (EMT) and ``Mesenchymal to Epithelial'' transitions (MET).  During the EMT process, a cell loses adhesion to neighbouring cells, and becomes more invasive and migratory.  It is worth noting that both EMTs and METs are part of normal developmental processes such as embryogenesis and tissue healing. Nevertheless, they are of one of main mechanisms of tumor metastasis. After undergoing EMTs, cancerous cells travel through the blood as Circulating Tumor Cells  (CTCs). These CTCs settle in other organs by undergoing METs, and they subsequently multiply, thus giving rise to metastatic tumors~\cite{thiery09,lambert17}.  Because of their key role in cancer, the identification of the GRNs enabling EMTs/METs has been a focus of a large research effort. Transcription factors (TFs) such as SNAIL, SLUG, TWIST, and ZEB1 have been studied in great detail \cite{decraene13,lamouille14}.  Cellular signals such as p54, Notch, EGF, Wnt, HIF-1$\alpha$, and others can induce EMTs/METs \cite{thiery09}. These signals act on a core four-component network that involves the upstream SNAIL/miR-34 circuit and the downstream ZEB1/miR-200 circuit \cite{levine13,kolch15}. The conceptual organization for such a circuit is depicted in Figure~\ref{f.circuit}. Each pair reproduces the standard toggle switch architecture, i.e., mutual inhibition. However, this design differs by having a mixed inhibition mechanism: the TFs ZEB1 and SNAIL inhibit the miRNAs ($\mu_{34},\mu_{200}$) at the transcriptional level, while $\mu_{34},\mu_{200}$ inhibit SNAIL and ZEB1, respectively, at the translational level. Therefore, such circuits are known as \emph{chimeric} circuits~\cite{levine13}.

One of the contributions of this work is to translate the conceptual diagram in Figure~\ref{f.circuit} into a precise mathematical model.  Our model, while not completely novel, clarifies and provides explicit details of several features of models found in the literature.
More importantly, we analyze the long-term behavior of a reduced model obtained under a natural and biologically realistic time scale separation assumption. We prove a theorem guaranteeing ``almost-sure'' convergence of trajectories to steady states.
This paper also serves to motivate a powerful but relatively unknown theorem on singularly perturbed monotone systems.  A central and well-known result for monotone systems is Hirsch's Generic
Convergence Theorem 
\cite{Hirsch83,Hirsch85,Hirsch88,Hirsch-Smith,Smith_book}, which guarantees that almost every bounded solution of
a strongly monotone system converges to the set of steady states. This theory has been widely applied to biochemical systems \cite{angeli04,sontag07,angeli10}.  However, many biological models are not monotone.
For example, monotonicity with respect to orthant cones rules out negative feedback loops, which are key components of homeostasis and adaptive systems. Indeed, the core EMT/MET network that is the focus of this paper has a negative feedback loop between $S$ and $m_S$ (see Figure~\ref{f.circuit}).  Nevertheless, we can take advantage of the fact that transcription happens at a faster time-scale than translation. Moreover, despite the fact that miRNAs and mRNAs belong to the same class of biochemical molecules, comparison of their half-lives reveals that mRNAs have much shorter half-lives than miRNAs and proteins \cite{sharova08,guo15}.   Hence, the negative loop in the core EMT/MET network is a ``fast'' loop. Intuitively, negative loops that act at a
comparatively fast time scale should not affect the main characteristics
of monotone behavior.  This has been rigorously studied in~\cite{wang:JNS07} (see the Ph.D.\ thesis~\cite{wang08} for more details), using tools from geometric singular perturbation theory. 

The paper is organized as follows. In Section II we model the core EMT/MET network in detail, and derive a reduced model.  Section III reviews several basic definitions and theorems about monotone systems, and in Section IV we analyze the reduced model theoretically and we review results that are can be used to extend convergence from reduced to full systems.

\section{Modeling of the EMT/MET system}
A mathematical model for the EMT/MET system has been presented in \cite{lu13,levine13} based on several assumptions that include detailed balance. Here, we drop such assumptions, and develop a more ``mechanistic'' model via the framework of Chemical Reaction Networks (CRN).  We review the notation briefly \cite{erdi,MA_PLOS20}. A CRN is a set of species $\mathscr S=\{Z_1,..,Z_n\}$ and a set of reactions $\mathscr R=\{\rm R_1,...,\rm R_\nu\}$. A reaction $\rm R_j$ can be written as: $\sum_{i=1}^n \alpha_{ij} Z_i \to \sum_{i=1}^n \beta_{ij} Z_j $.  

A stoichiometry matrix $\Gamma \in \mathbb R^{n \times \nu}$ is defined elementwise as $[\Gamma]_{ij}=\beta_{ij}-\alpha_{ij}$. The reactions are associated with a rate function $R: \mathbb R_{\ge 0}^n \to \mathbb R_{\ge 0}^\nu$.  We assume that $R$ takes the form of Mass-Action kinetics: $ R_j(z) = \prod_{i=1}^n k_j z_j^{\alpha_{ij}}$, where $k_j$ is the kinetic constant.
Let  $z(t) \in \mathbb R_{\ge 0}^n$ be  the vector of species concentrations at time $t$. 
The associated ODE can be written as: $\dot z = \Gamma R(z)$.

We model  GRNs as CRNs via the central dogma of molecular biology (see \cite{MA_PLOS} for a detailed framework). Each gene is associated with promoter states, an mRNA state, and a protein state. As mentioned in the introduction, we can assume that the promoter kinetics and the mRNA kinetics are fast. In this section, we will derive the dynamics of the fast and slow systems. We will show that the model can be reduced from \textit{24 dimensions} to \textit{four dimensions}. The slow states are $S,Z, \mu_{200}, \mu_{34}$.  
We will start with promoter kinetics.

\subsection{Promoter dynamics}
For each gene  {$j$}, we denote the promoter states  by species of the form $D_i^j$. The superscript denotes the gene, while the subscript denotes the occupancy of the binding sites. Let $D^{34}, D^{200}, D^{S}, D^{Z}$ be the unbound promoters for $\mu_{34},\mu_{200}$, SNAIL, and ZEB1, respectively.  For instance, $D_s^{34}$ denotes $S$ binding to the promoter of $\mu_{34}$, while $D_{sz}^{34}$ denotes both $S,Z$ binding to the promoter of $\mu_{34}$. 

Hence, the CRN  describing promoter dynamics for  the EMT/MET network (Figure \ref{f.circuit}) can be written as:
\begin{align*} 
2S+ D^{Z} & \xrightleftharpoons[ \alpha_{-s}]{ \tfrac{\alpha_{s}}2}    D_s^{Z}, \ \  2Z+ D^{Z}   \xrightleftharpoons[ \alpha_{-z}]{ \tfrac{\alpha_{z}}2}    D_z^{Z},  \ 2S+ D_z^{Z}    \xrightleftharpoons[ \alpha_{-s}]{ \tfrac{\alpha_{s}}2}    D_{sz}^{Z}, \ \  2Z+ D_s^{Z}   \xrightleftharpoons[ \alpha_{-z}]{ \tfrac{\alpha_{z}}2}    D_{sz}^{Z}, \\
2S+ D^{34} & \xrightleftharpoons[ \alpha_{-s}]{ \tfrac{\alpha_{s}}2}    D_s^{34},  \  2Z+ D^{34}   \xrightleftharpoons[ \alpha_{-z}]{ \tfrac{\alpha_{z}}2}    D_z^{34},  \ 2S+ D_z^{34}    \xrightleftharpoons[ \alpha_{-s}]{ \tfrac{\alpha_{s}}2}    D_{sz}^{34},  \  2Z+ D_s^{34}   \xrightleftharpoons[ \alpha_{-z}]{ \tfrac{\alpha_{z}}2}    D_{sz}^{34}, \\
2S+ D^{200} & \xrightleftharpoons[ \alpha_{-s}]{ \alpha_{s}/2}    D_s^{200}, \quad 2Z+ D^{200}   \xrightleftharpoons[ \alpha_{-z}]{ \alpha_{z}/2}    D_z^{200},   2S+ D_z^{200}    \xrightleftharpoons[ \alpha_{-s}]{ \alpha_{s}/2}    D_{sz}^{200},\\
2Z+ D_s^{200}  & \xrightleftharpoons[ \alpha_{-z}]{ \alpha_{z}/2}    D_{sz}^{200}, \ \ 2S+ D^{S}   \xrightleftharpoons[ \alpha_{-s}]{ \alpha_{s}/2}    D_s^{S}, 
\end{align*}
The concentration of a species $Y$ will be denoted, if no other notation is used, by $[Y]$.
The reaction structure implies that we have the conservation law $\sum_{i} [D_i^j](t)=E_j$ for each gene, where $E_j, j\in\{S,Z,34,200\}$ is the total concentration available  in the medium. Hence,  $E_j$ stays constant during the course of the reaction.   Note that  each TF is assumed to bind to its promoter as a dimer. The same analysis can be repeated if the TF binds as an $n$-mer for some integer $n \ge 1$. 

Since binding/unbinding kinetics are usually fast \cite{alon06}, we will approximate all the promoter states with their quasi steady state (QSS) approximations.  We are interested in deriving the expression for the \emph{active} promoter states.  Since both $S,Z$ repress $\mu_{34},\mu_{200}$, then the active states are the unbound promoter states $D^{34}, D^{200}$, respectively.  {The promoter $D^S$ is the active state for SNAIL since it is a self-repressing gene. Finally, $ D_{sz}^{Z}$ is the active state for ZEB1 because it is activated by both SNAIL and self-binding}.

We will consider ZEB1 as an example. Let $d_z(t):=[ [D^{Z}],[D_s^{Z}], [D_z^{Z}], [D_{sz}^{Z}]]^T(t)$. Using the reactions above, we can write the ODE for the promoters of $Z$ to get: 
\begin{align}\label{D}
{\dot d_z}  = P_z(s,z) d_z :=\begin{bmatrix}  -\alpha_ss^2- \alpha_z z^2 & \alpha_{-s} & \alpha_{-z} & 0 \\ \alpha_s s^2 & -\alpha_{-s}-z^2\alpha_z & 0 & \alpha_{-z}  \\ \alpha_z z^2 & 0 & -\alpha_{-z}-s^2\alpha_s  &  \alpha_{-s}  \\  0 & z^2 \alpha_z & s^2 \alpha_s & -\alpha_{-s}-\alpha_{-z} \end{bmatrix}d_z.
\end{align}
Note that $s,z$ are slow variables. 
Setting the derivatives to zero and substituting the conservation law  we get:
\[ [D_{sz}^{Z}]_{qss}=\frac{ {E_{Z}}s^2 z^2}{(s^2+A_s)(z^2+A_z)}, \]
where $A_s:=\alpha_{-s}/\alpha_s, A_z:=\alpha_{-z}/\alpha_z$.
Similarly, we can define the matrices $P_s(s,z),P_{34}(s,z),P_{200}(s,z)$. Hence, we get:
\[ [D^{200}]_{qss}=\frac{E_{200} {A_s A_z}}{(s^2+A_s)(z^2+A_z)}, [D^{S}]_{qss}=\frac{E_{S} {A_s} }{(s^2+A_s)}, [D^{34}]_{qss}=\frac{E_{34}  {A_s A_z} }{(s^2+A_s)(z^2+A_z)}.  \]

\subsection{RNA dynamics}
Translational inhibition by miRNAs is achieved by an RNA-induced Silencing Complex (RISC). RISC binds to the target mRNA and degrades it via the Argonaute protein \cite{ender10}.  Here, we assume a simple model in which miRNA binds to the target mRNA to inhibit translation and activate degradation.  In general, multiple miRNAs can bind to a single mRNA. Here we assume that each mRNA can have up to two binding sites. For each additional miRNA binding, translation is inhibited and degradation is accelerated.

The CRN for the transcription of SNAIL and ZEB1 can be written as follows:
\begin{align*} 
D^{S} & \xrightarrow{ \gamma_{s}}    D^{S}+M_s,  \ \  D_{sz}^{Z}  \xrightarrow{ \gamma_{z}}    D^{Z}+M_z, \ \ M_s \xrightarrow{ \beta_{s}} \emptyset, \ \ M_z \xrightarrow{ \beta_{z}} \emptyset ,
\end{align*}
where $M_s,M_z$ are the species denoting the mRNAs of SNAIL and ZEB1, respectively.
The CRN for the miRNA-mRNA reaction (with two binding sites) can be written as follows:
\begin{align*}
\mu_{34} + M_S  &  \xrightleftharpoons[ c_{-s}]{ c_s} M_{S\mu} \xrightarrow{\beta_{s1}} \mu_{34}, \ \ \mu_{34} + M_{S\mu}     \xrightleftharpoons[ c_{-s}]{ c_s} M_{S\mu_2} \xrightarrow{\beta_{s2}} 2\mu_{34},  \\
\mu_{200} + M_Z  &  \xrightleftharpoons[ c_{-z}]{ c_z} M_{Z\mu} \xrightarrow{\beta_{z1}} \mu_{200}, \ \ \mu_{200} + M_{Z\mu}     \xrightleftharpoons[ c_{-z}]{ c_z} M_{Z\mu_2} \xrightarrow{\beta_{z2}} 2\mu_{200}, 
\end{align*}

Since miRNAs actively degrade mRNAs, we assume that:
\begin{equation}\label{bineq} \beta_s < \beta_{s1} < \beta_{s2}, \quad  \beta_z < \beta_{z1} < \beta_{z2} . \end{equation}
Remembering that $\mu_{200}$, and  $z$ are the slow variables, we can write the ODE for ZEB1 translational dynamics as follows (with $m_z(t):= [  [M_z] , [M_{z\mu}] , [M_{z\mu_2}]]^T(t)$):
\begin{align}\label{M}
\dot m_z   & =Q_z(\mu_{200}) m_z + b_z  {D_{sz}^Z} \\ \nonumber &:=\begin{bmatrix} -c_z\mu_{200}-\beta_z &  c_{-z} & 0 \\ c_z\mu_{200} & -\beta_{z1}-c_z \mu_{200}-c_{-z} & c_{-z}  \\ 0 & c_z \mu_{200} & -c_{-z}-\beta_{z2} \end{bmatrix} m_z+\begin{bmatrix}   \gamma_z \\ 0 \\ 0
\end{bmatrix}  {D_{sz}^Z}
\end{align}
Since \eqref{M}  is a linear system in $\mu_{200}$,  the quasi-steady state can be solved easily by matrix inversion. Substituting the QSS for $[D^Z]$ we get:
\begin{equation}\label{mz} m_{z,qss} =  \frac{\gamma_z E_Z s^2 z^2}{(s^2+A_s)(z^2+A_z) } \begin{bmatrix}\beta_{z2} c_z \mu_{200}+e_{z1}\\ e_{z2}\mu_{200}  \\ c_z^2 \mu_{200}^2 \end{bmatrix}\frac 1{\beta_{z2} c_z^2 \mu_{200}^2 + e_{z3}c_z\mu_{200}+e_{z1}\beta_z},\end{equation}
where $e_{z1}:=\beta_{z1}\beta_{z2}+(\beta_{z1}+\beta_{z2})c_{-z}+c_{-z}^2, e_{z2}:=c_z(\beta_{z2}+c_{-z}), e_{z3}:= \beta_z \beta_{z2} + \beta_{z1} \beta_{z2}+\beta_{z1} c_{-z}$.
The dynamics of $m_s$ can be derived similarly and we get:
\begin{equation}\label{ms} m_{s,qss} =  \frac{\gamma_s E_S  {A_s} }{(s^2+A_s)  } \begin{bmatrix}\beta_{s2} c_s \mu_{34}+e_{s1}\\ e_{s2}\mu_{34}  \\ c_s^2 \mu_{34}^2 \end{bmatrix}\frac 1{\beta_{s2} c_s^2 \mu_{34}^2 + e_{s3}c_s\mu_{34}+e_{s1}\beta_s},\end{equation}
where $e_{s1}:=\beta_{s1}\beta_{s2}+(\beta_{s1}+\beta_{s2})c_{-s}+c_{-s}^2, e_{s2}:=c_s(\beta_{s2}+c_{-s}), e_{s3}:= \beta_s \beta_{s2} + \beta_{s1} \beta_{s2}+\beta_{s1} c_{-s}$.
\subsection{Slow Dynamics}
The complete reaction model requires modeling the production of the proteins and the miRNAs. The CRN can be written as follows:
\begin{align*} D^{34}  & \xrightarrow{\varepsilon \beta_{34}}    D^{34}+\mu_{34},  \ \ \mu_{34} \xrightarrow{ \varepsilon\beta_{-34}} \emptyset, \\
D^{200} & \xrightarrow{\varepsilon \beta_{200}}    D^{200}+\mu_{200}, \ \ \mu_{200} \xrightarrow{\varepsilon \beta_{-200}} \emptyset, \\
M_s  &\xrightarrow{\varepsilon k_{s}} M_s + S, \ \ M_{s\mu}  \xrightarrow{\varepsilon k_{s1}} M_{s\mu} + S  , \ \ M_{s\mu_2}  \xrightarrow{\varepsilon k_{s2}} M_{s\mu_2} + S, \ \ S  \xrightarrow{\varepsilon k_{-s}} \emptyset, \\
M_z  &\xrightarrow{\varepsilon k_{z}} M_z + Z, \ \ M_{z\mu}  \xrightarrow{\varepsilon k_{z1}} M_{z\mu} + Z  , \ \ M_{z\mu_2}  \xrightarrow{\varepsilon k_{z2}} M_{z\mu_2} + Z ,\ \ Z \xrightarrow{\varepsilon k_{-z}} \emptyset.
\end{align*}

The kinetic  constants are multiplied by $\varepsilon$ to emphasize that the corresponding reactions  are slow.
Note that since miRNA inhibits translation, the  mRNA-miRNA complexes have lower translation rates than raw mRNAs. Hence, we have: 
\begin{equation}\label{kineq}k_s > k_{s1} > k_{s2}, \quad k_z > k_{z1} > k_{z2}. \end{equation}

The model above assumes that $Z$ is produced only when both $Z$, and $S$ are bound to ZEB1's promoter. But this is not realistic, since constitutive transcription is always present  at low levels. Hence, we model this by a reaction $\emptyset \xrightarrow{\varepsilon\delta_z} Z $ with $\delta_z$ small. This will help us also show generic convergence for the reduced system in \S 4.

After substituting the QSSs for the fast variables in the slow system, all the terms corresponding to promoter and mRNA reactions vanish. Hence, the corresponding ODE can be written as follows:
\begin{align} \nonumber
\dot  \mu_{34}&= \beta_{34}[D^{34}]-\beta_{34} \mu_{34} = \frac{E_{34}  {A_z A_s}\beta_{34}}{(s^2+A_s)(z^2+A_z)}-\beta_{34} \mu_{34} \\ \label{slow}
\dot s&= [k_s \ k_{s1} \ k_{s2}] m_s-k_{-s} s = [k_s \ k_{s1} \ k_{s2}] m_{s,qss}-k_{-s} s \\\nonumber
\dot  \mu_{200}&= \beta_{200}[D^{200}]-\beta_{200} \mu_{200} = \frac{E_{200}\beta_{200} {A_z A_s} }{(s^2+A_s)(z^2+A_z)}-\beta_{200} \mu_{200} \\\nonumber
\dot z&= \delta_z+ [k_z \ k_{z1} \ k_{z2}] m_z-k_{-z} z = \delta_z+[k_z \ k_{z1} \ k_{z2}] m_{z,qss}-k_{-z} z.
\end{align}

where $m_{z,qss},m_{s,qss}$ are given by \eqref{mz},\eqref{ms}, respectively.

\section{Monotone systems and singular perturbations} 

\subsection{Monotone systems}
In this section, we review basic definitions and results regarding
monotone systems. We base our discussion on \cite{Hirsch85,Smith_book,wang08}.

 	A nonempty, closed set $C \subset \mathbb R^N$ is said to be a \textit{cone} if  $C+C \subset C$,  $\alpha C \subset C$ for all $\alpha>0$,  and $C \cap \,(-C)= \{0\}$.
For each cone $C$, a partial order on $\mathbb R^N$ can be associated. For any $x,y
\in \mathbb R^N$, we define:
\begin{align*}
x \geq y & \Leftrightarrow x-y \in C \\
x > y & \Leftrightarrow x-y \in C, x \not= y.
\end{align*}
When  $C^\circ$ is not empty, we can define $ x \gg y \Leftrightarrow x-y \in  C^\circ$. 
 
 In this paper, we only consider cones which are \textit{orthants} of $\mathbb R^N$.   In order to identify the various orthants, let $\sigma \in \{\pm 1 \}^N$ and let \[C_\sigma=\{z \in \mathbb R^n | \sigma_i z_i \ge 0, 1 \le i \le n \}\]
 be the corresponding orthant cone.  Let $\preceq_\sigma$ be the corresponding partial order.

A set $W \subseteq \mathbb R^N$ is said to be $p$-convex, if $W$ contains the  line  joining $x$ and $y$ whenever $x \preceq y$, $x,y \in W$.
Hence, equipped with a partial ordering on a $p$-convex and \emph{open} set $W$  we study the ordinary differential equation: 
\begin{eqnarray}
\frac{d z}{dt}=F(z), \label{eqn:def}
\end{eqnarray}
where $F:W \rightarrow \mathbb R^N$ is a $C^1$ vector field. We assume that the system is  forward invariant  with respect to  $W$.  In our application in this paper we will use $W=\mathbb R_+^N$, the \emph{open} positive orthant.

We are interested in a special class of equations which preserve the partial order along all trajectories.  

\begin{definition}
 {	The flow $\phi_t$ of (\ref{eqn:def}) is said to have   positive derivatives on a set $W \subseteq \mathbb R^N$, if $\left [\tfrac{\partial}{\partial z} \phi_t( z)\right]x \in C^\circ$ for all $x \in C \setminus \{0\},z \in W$, and $t  > 0$. }
\end{definition}

\begin{definition}
	The system \eqref{eqn:def}  is called \textit{monotone} (resp. strongly monotone) on a set $W \subseteq \mathbb R^N$  if for all 
	$t > 0$ and all $z_1,z_2 \in W$,
	\begin{align*}
	z_1 \geq z_2 \Rightarrow  \phi_t(z_1) \geq \phi_t(z_2)  
	\  (\mbox{resp.} \  \phi_t(z_1) \gg \phi_t(z_2) \ \mbox{when} \ z_1 \not= z_2).
	\end{align*}
\end{definition}

Establishing that a flow  has  positive derivatives can be performed by verifying the irreducibility of the Jacobian as the following theorem states:
\begin{proposition}
{	Assume that \eqref{eqn:def} is monotone with respect to an orthant cone $C$.  If $ \frac{\partial F}{\partial z}( z) $ is  irreducible for all $z \in W$   then the flow $\phi_t$ of (\ref{eqn:def}) has  positive derivatives on the set $W \subseteq \mathbb R^N$. %
 }
\end{proposition}

The following result can be interpreted as saying that having positive derivatives is an "infinitesimal" version of strong monotonicity.

\begin{proposition}
{	Let $W\subset \mathbb R^N$ be  $p$-convex and open. If  the  flow $\phi_t$ has  positive derivatives in $W$, then the associated ODE is strongly monotone on $W$.}
\end{proposition}

Hence, we have the following corollary: 
\begin{corollary}
Let \eqref{eqn:def} and $W$ be given as above. Assume that \eqref{eqn:def} is monotone with respect to an orthant cone $C$.  If $ \frac{\partial F}{\partial z}( z) $ is  irreducible for all $z \in W$, then the system \eqref{eqn:def} is strongly monotone with respect to $C$ on $W$.
\end{corollary}

This is a rephrasing of the well-known Hirsch Generic Convergence Theorem:

\begin{theorem}
	\label{lemma:Hirsch_conv}
	 {Assume that  \eqref{eqn:def} is strongly monotone on a $p$-convex open set $W \subseteq \mathbb R^N$. Let $W^c \subseteq W$ be defined as the subset of points whose forward orbit has compact closure in $W$. If the set of equilibria is totally disconnected, then the forward trajectory starting from almost every point in $W^c$ converges to an equilibrium.}
\end{theorem}

By Corollary 1, it is sufficient to check monotonicity and irreducibility to establish generic convergence.

 \subsection{Graphical characterization of monotonicity}
Monotonicity with respect to orthant cones can be characterized via \emph{Kamke's conditions}. We review the relevant material from \cite{Smith_book,sontag07}.  Let $\sigma \in \{\pm 1 \}^N$ and let $C_\sigma$ be the corresponding orthant as defined above. Let $\Sigma=\mbox{diag}(\sigma)$, i.e., a diagonal matrix with $\sigma$ as the diagonal. Then, the following holds:
\begin{theorem}(Kamke's conditions) \label{th.kamke} Let    \eqref{eqn:def} be given and let $\phi_t$ be the associate flow. Let $J=\frac{\partial F}{\partial z}$ be the corresponding Jacobian.  Let  $W \subseteq \mathbb R^N$ be a $p$-convex open set. Assume that there exists $\sigma \in \{\pm 1 \}^N$ such that $\Sigma J \Sigma$ is Metzler on $W$ (i.e., all non-diagonal entries are non-negative). Then the corresponding flow is monotone on $W$ with respect to the partial order $\preceq_\sigma$. %
	\end{theorem}

The last proposition gives a useful characterization for monotonicity with respect to orthant cones; however, it requires checking $2^N$ possible sign combinations. Alternatively,  a simple graphical criteria   can be stated for a graph derived from the Jacobian. Informally, it states that the system is orthant monotone if \textit{every loop has a net positive sign}.  We state it more formally next. 

We say that a Jacobian $J$ is \textit{sign-stable} on a set $W$ if for each $i\ne j$, $\mbox{sgn}(J_{ij})$ is constant on $W$.  Define a signed \textit{directed} graph $G$ with vertices $\{1,...,n\}$. There is an edge connecting vertices $i$ to $j$ if  the partial derivative $J_{ij}$ does not vanish identically. The sign of the edge is equal to sign of $J_{ij}$.  A \emph{loop} is any sequence of edges (without regard to direction) that does not traverse a vertex twice and it starts and ends with the same vertex. The \emph{sign} of a loop is the product of the signs of the constituent edges. It is worth noting that diagonal entries, i.e., ``self-loops'', have no effect on the validity of the results.

\begin{theorem}(Positive Loop Property) \label{th.loop} Let \eqref{eqn:def} be given. Let $J=\frac{\partial F}{\partial z}$ be the corresponding Jacobian. Assume $J$ is sign-stable.  Let  $W \subseteq \mathbb R^N$ be a $p$-convex open set. Let $G$ be the graph defined above. If every loop has a positive sign, then there exists $\sigma \in \{\pm 1 \}^N$ such that $\Sigma J \Sigma$ is Metzler on $W$ (i.e, all non-diagonal entries are non-negative).
\end{theorem}

 \subsection{Singular perturbation model reduction}
The process of mathematical modeling of physical processes involves usually reduction of the dynamics of fast variables.  For instance, self-loops, i.e., terms corresponding to the diagonal entries of the Jacobian $J$,  are often approximations of fast dynamics. Hence, it is intuitive to expect that the theorems in the previous section hold for \emph{sufficiently fast} negative self-loops. 

Hence, we consider a system in the 
 singularly perturbed form: 
\begin{align}
\label{eqn:slow_0}
\frac{dx}{dt}&= f_0(x,y,\varepsilon) \\
\varepsilon \frac{dy}{dt}&= g_0(x,y,\varepsilon),\notag
\end{align}
where $f_0:\mathbb R_+^n \times \mathbb R_+^m \times [0,\bar\varepsilon]  \to \mathbb R_+^n, g_0: \mathbb R_+^n \times \mathbb R_+^m \times [0,\bar\varepsilon] \to \mathbb R_{+}^m  $ are smooth bounded functions and $\bar\varepsilon >0$ is fixed. Furthermore, we assume that the equilibrium set is totally disconnected for all $\varepsilon < \bar\varepsilon$.
These assumptions are  automatically satisfied in our case since Mass-Action kinetics give rise to polynomial systems. 

For $0<\varepsilon\ll1$, the dynamics of $x$ are much
slower than $y$. 
If $\varepsilon\not=0$, we can change the time scale to
$\tau=t/\varepsilon$, and study the equivalent form:
\begin{align}
\label{eqn:f_xy}
\frac{dx}{d\tau}&=\varepsilon f_0(x,y,\varepsilon) \\
\frac{dy}{d\tau}&= g_0(x,y,\varepsilon). \notag
\end{align}

Model reduction via singular perturbations requires solving the fast system at a quasi-steady state, hence we assume that there exists a smooth bounded function 
\[
m_0:\mathbb R_m^+ \rightarrow \mathbb R_n^+
\]
such that $g_0(x,m_0(x),0)=0$ for all $x \in \mathbb R_+^n$.  From the previous section it can be seen that $m_0$ exists and is unique as we have derived all the QSS expressions uniquely.

For simplicity, let  $z=y-m_0(x)$. Hence, the fast system \eqref{eqn:f_xy} can be written as:
\begin{align}
\label{eqn:f_xz}
\frac{dx}{d\tau}&=\varepsilon f_1(x,z,\varepsilon)  \\
\frac{dz}{d\tau}&= g_1(x,z,\varepsilon), \notag
\end{align}
where
\begin{align*}
f_1(x,z,\varepsilon)&=f_0(x,z+m_0(x),\varepsilon),\\
g_1(x,z,\varepsilon)&=g_0(x,z+m_0(x),\varepsilon)-\varepsilon [\tfrac{\partial}{\partial x} m_0(x)] f_1(x,z,\varepsilon).
\end{align*}
When $\varepsilon=0$, the system \eqref{eqn:f_xz} becomes
\begin{equation}
\label{eqn:f_z}
\frac{dz}{d\tau}=g_1(x,z,0), \ \ x(\tau) \equiv x_0 \in W_x.
\end{equation}

Verifying the stability of the fast system is a standard and an intuitive requirement for singular perturbation methods as in the original Tikhonov theorem \cite{tikhonov52}.  Hence we need to check that:
\begin{enumerate}
	\item    the equilibrium $z=0$ of \eqref{eqn:f_z} is globally asymptotically stable on $\{z \,|\,z+m_0(x_0) \in \mathbb R_+^m\}$ for all $x_0 \in \mathbb R_n^+$. 
\item all eigenvalues of the Jacobian  $\tfrac{\partial }{\partial y} g_0(x,m_0(x),0)$  have negative real parts for every $x \in \mathbb R_n^+$.
\end{enumerate} We study this in the next section.

   \section{Analysis of the core EMT/MET network}
   In our analysis in the  section 2  we have considered fast and slow reactions separately. We will use now the notation from \S 3.  Let $x(t)=[\mu_{34},s,\mu_{200},z]^T(t)$ be the state of the slow system, and let $y(t)=[d_s^T,d_z^T,d_{34}^T,d_{200}^T,m_s^T,m_z^T](t)$ be the state of the fast system.  
   Hence, the full system can be written in the form \eqref{eqn:slow_0}. We denote the reduced system \eqref{slow} as $\dot x = G(x)$. 
   \subsection{Stability of the fast dynamics}
     In this subsection, we want to study global asymptotic stability of the fast system, and to show that its Jacobian is Hurwitz.

For a fixed slow variable $x$,  the fast dynamics of ZEB1, SNAIL, miRNA-34 , miRNA-{200} are decoupled from each other as can be seen from the CRNs introduced before. Furthermore, all the systems are linear. Let us analyze ZEB1 as an example. From \eqref{D},\eqref{M} we get
\[ \begin{bmatrix} \dot d_z \\ \dot m_z\end{bmatrix}= \begin{bmatrix} P_z(s,z) & 0 \\  {B_z} & Q_z(s,z) \end{bmatrix}\begin{bmatrix}  d_z \\  m_z\end{bmatrix}, \]
 {where $B_z:=[ O_{3 \times 3}, b_z]$ and $O_{3 \times 3} \in \mathbb R^{3 \times 3 }$ is a zero matrix}.
The dynamics of $d_z$ are decoupled from $m_z$. Since $P_z$ is an irreducible Metzler matrix with principal  eigenvalue of 0, Perron-Frobenius Theorem \cite{berman94} implies that all other eigenvalues are strictly negative. In fact, they can be computed as $\{-\alpha_s s^2-\alpha_{-s},-\alpha_z z^2-\alpha_{-z}, -\alpha_s s^2-\alpha_{-s}-\alpha_z z^2-\alpha_{-z}\}$. Since the total number of promoters is conserved, we can reduce  the dimension of the ODE by one.  It follows that the reduced promoter dynamics are globally asymptotically stable. The same argument applies to the promoters of $Z, \mu_{34}, \mu_{200}$. 

We turn to the dynamics of $m_z$. Due to the block triangular structure we only need to study the eigenvalues of $Q_z(s,z)$. This matrix is also Metzler and the principal eigenvalue can be shown to be negative. Hence, it is Hurwitz. This can be proven alternatively with a linear Lyapunov function $V(m_z)=\mathbf 1^T m_z$. The time-derivative is $\dot V(m_z)= -[\beta_z \ \beta_{z1} \ \beta_{z2}] m_z <0$ for all $m_z \ne 0$.  The same argument applies to the translational dynamic of SNAIL. 
Hence, the fast system is globally asymptotically stable and its  Jacobian is Hurwitz.  
\subsection{Monotonicity of the reduced system}
We study the reduced system \eqref{slow}. We will utilize Kamke's conditions as in Theorem \ref{th.kamke}. Hence, we compute the Jacobian $J$ for the slow system \eqref{slow}:

\[J=\begin{bmatrix} -\beta_{34} & -\frac{2E_{34} {A_s A_z}\beta_{34}s}{(s^2+A_s)^2(z^2+A_z)} & 0 & -\frac{2E_{34} {A_s A_z}\beta_{34}z}{(s^2+A_s)^2(z^2+A_z)} \\ J_{21} & J_{22} & 0 & 0 \\ 0 & -\frac{2E_{200} {A_s A_z}\beta_{200}s}{(s^2+A_s)^2(z^2+A_z)}&-\beta_{200} & -\frac{2E_{200} {A_s A_z}\beta_{200}z}{(s^2+A_s)^2(z^2+A_z)} \\ 0 & J_{42}& J_{43}& J_{44} \end{bmatrix},\]
where 
\[ J_{21}= -\frac{h_1(\mu_{34}) E_{s} {A_s}\gamma_s}{(s^2+A_s)(\beta_{ {s}2} c_ {s}^2 \mu_{ {34}}^2 + e_{ {s}3}c_ {s}\mu_{ {34}}+e_{ {s}1}\beta_ {s})^2},\]
where $h(\mu_{ {34}}):=c_{s} ( \beta_{s2} + c_{-s})^2 ( \beta_{s1} + c_{-s}) (\beta_{s1} k_{s}-\beta_{s} k_{s1} ) + 2 c_{s}^2 \mu_{34} (k_{s} \beta_{s2} - \beta_{s} k_{s2})  (c_{-s}+\beta_{s1}) (c_{-s}+\beta_{s2})+c_{s}^3 \mu_{34}^2 (\beta_{s2} (\beta_{s2} k_{s} - \beta_{s} k_{s2} )+ (\beta_{s2}  + c_{-s}) (\beta_{s2} k_{s1}- \beta_{s1} k_{s2}))$. Note that $h(\mu_{200})>0$ for all parameters if \eqref{bineq},\eqref{kineq} are satisfied. Hence $J_{21}<0$. Similarly, we can also calculate $J_{43}$ to show that it is negative whenever \eqref{bineq},\eqref{kineq} are satisfied.

Finally, using \eqref{mz} we compute $J_{42}$ as:
\[ J_{42}=\frac{2 E_Z \gamma_z A_s s z^2}{(s^2+A_s)^2(z^2+A_z) } \frac{ k_z(\beta_{z2} c_z \mu_{200}+e_{z1})+k_{z1} (e_{z2}\mu_{200} ) +k_{z2} c_z^2 \mu_{200}^2 }{\beta_{z2} c_z^2 \mu_{200}^2 + e_{z3}c_z\mu_{200}+e_{z1}\beta_z}>0. \]
\begin{remark} Conditions \eqref{bineq},\eqref{kineq} are stronger than we need. In fact, it can be seen from the expression of $h$ above that it is sufficient to have the protein production ratio of the raw mRNA   greater than the corresponding one for the miRNA-RNA complex.
	\end{remark}
\begin{figure}[t]
	\centering
	\includegraphics[width=0.5\textwidth]{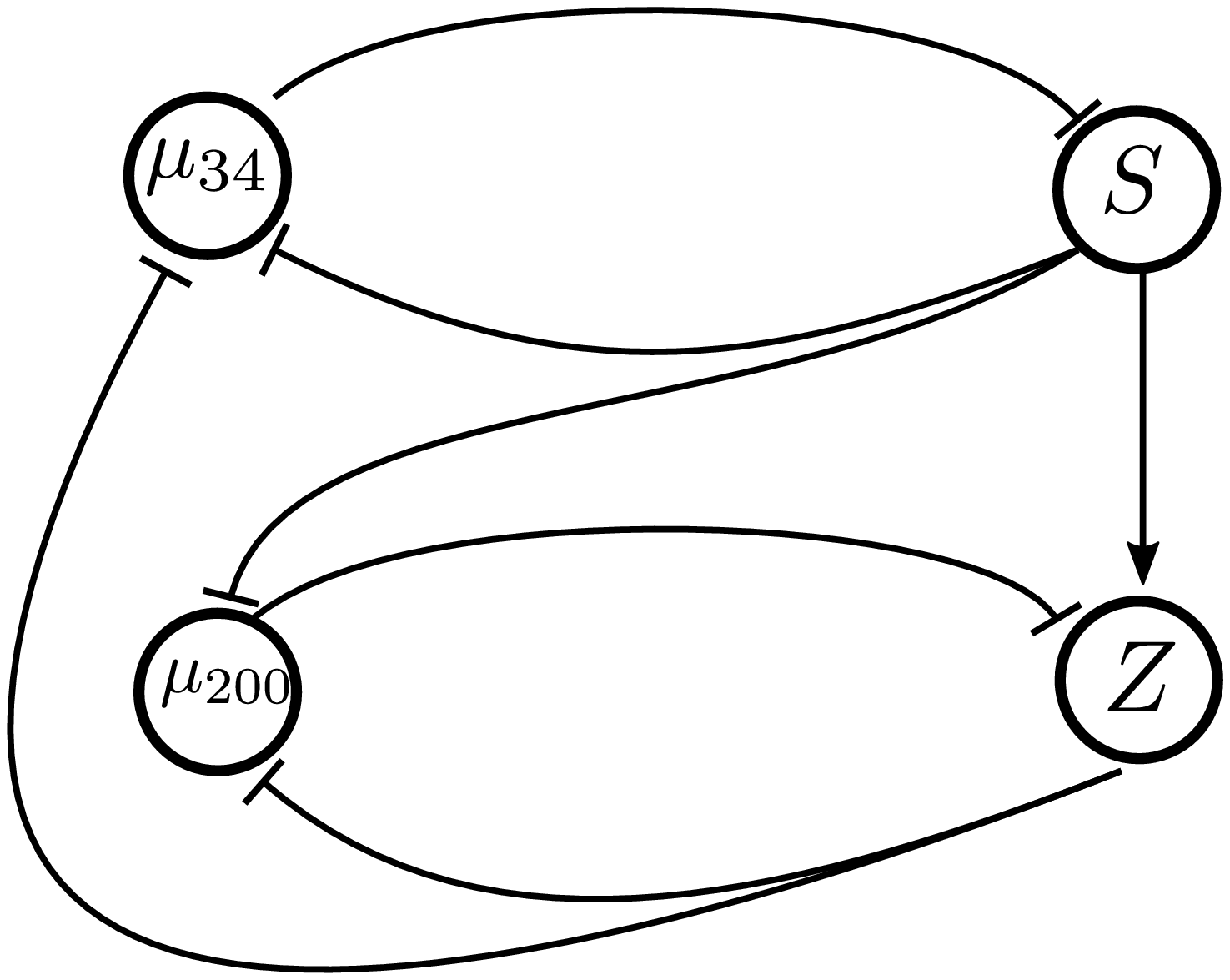}
	\caption{Graph of the reduced circuit. In the terminology of Theorem \ref{th.loop}, ``$\to$'' has a $+$ sign, and ``$\dashv$'' has a $-$ sign.} \label{f.circuit2}
\end{figure}
The signs of $J_{22}, J_{44}$ do not play a role since they correspond to self-loops. Therefore, we have the following sign pattern for the Jacobian under the conditions \eqref{bineq},\eqref{kineq}:
\[\begin{bmatrix} * & - & 0 & - \\ - & * & 0 & 0 \\ 0 & - & * & - \\ 0 & + & - & * \end{bmatrix},\]
which can be represented via the graph depicted in Figure \ref{f.circuit2}. It can be easily seen that every loop is positive. Hence, by Theorem 3 there exists $\sigma \in \{\pm 1\}^N$ such that $\Sigma J \Sigma$ is Metzler.      Therefore, by Theorem 2  we get that the reduced system \eqref{slow} is   monotone on $W=\mathbb R_{+}^4$. Monotonicity holds with respect to the orthant cone specified by   $\sigma=[-1,1,-1,1]^T$. (see \S 3.2) 

It can be verified that the Jacobian is irreducible on the \textit{open} positive orthant $\mathbb R_+^n$. Hence, by Corollary 1 the reduced system is strongly monotone on $\mathbb R_+^n$.

Fix an initial condition $x_0 \in \mathbb R_{+}^n$, and let $x(t):=\varphi(t;x_0)$ be the corresponding solution of the slow system. In order to infer generic convergence to the set of equilibria, we need the strong order preserving property to hold on the $\omega$-limit set of the solution. Hence, we need to show that $\omega(\varphi(t;x_0))\cap \partial \mathbb R_+^n = \emptyset$. In other words, we need to show that the slow system is \textit{persistent}.  This can be shown as follows. Recall that the slow system \eqref{slow} is given as $\dot x=G(x)$. It can be verified from \eqref{slow} that   $\forall i, G_i(z)>  0$ whenever $z_i=0$. Hence, there exists $\eta>0$ such that $\dot x_i(t)>0$ whenever $x_i(t) \in [0,\eta)$.  
This implies that the \textit{boundary has no equilibria} and is \textit{repelling} (i.e., the vector field is pointing away from the boundary). We proceed by seeking  contradiction. W.l.o.g,  assume that $\exists x^* \in \omega(\varphi(t;x_0))\cap \partial \mathbb R_+^n$ with $x_i^*=0$ (the $i$-coordinate). Hence, there exists a sequence $\{t_k\}_{k=1}^\infty$ such that $x_i(t_k)>0$ and $x_i(t_k) \to x_i^*= 0$. Therefore, for each right neighborhood $\mathcal N$ of 0 there exists $t^*$ for which $x_i(t^*) \in \mathcal N$ and $\dot x_i(t^*)<0$, which is a contradiction. Hence, the reduced system is persistent.

Therefore, we state the following theorem.
	
	\begin{theorem}
	Consider the reduced system \eqref{slow}. Let $W^c \subset \mathbb R_+^4$ be the subset of points whose forward orbit has a compact closure in $\mathbb R_+^4$. Then, the forward trajectory starting from almost every point in $W^c$ converges to an equilibrium.
	\end{theorem}

   \subsection{Remarks on generic convergence for singularly perturbed monotone systems}

\begin{figure}[ht]
	\centering \includegraphics[scale=0.4]{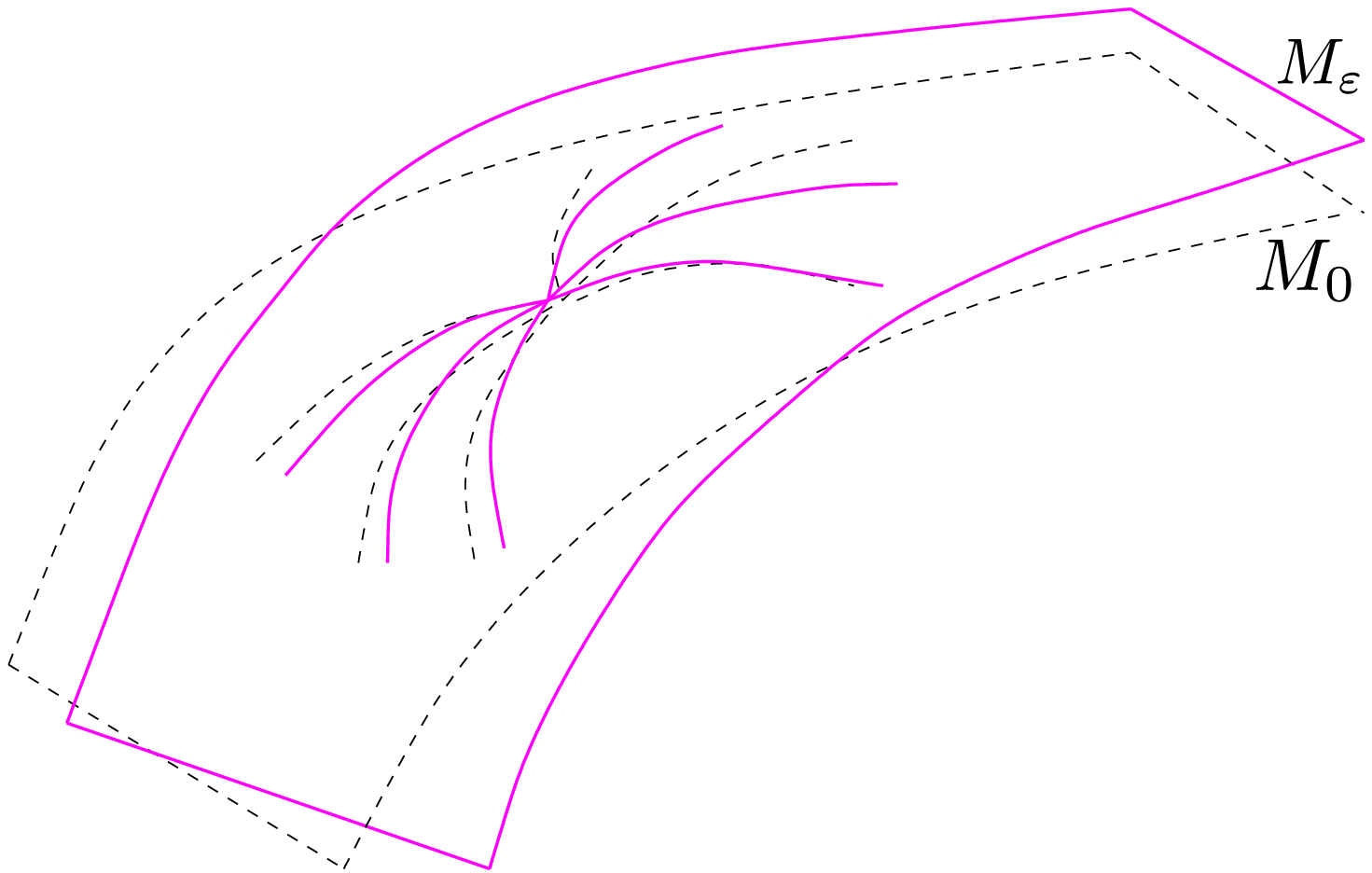}
	\caption{Illustration of the manifolds $M_0$ and $M_\varepsilon$. The figure shows two key properties of  $M_\varepsilon$. First, $M_\varepsilon$ is close to $M_0$. Second, the trajectories on $M_\varepsilon$ converge to 
		steady-states if those on $M_0$ do.}
	\label{fig:M_e}
\end{figure}

An interesting general question for singularly perturbed systems is as follows.  Assuming that the slow system is strongly monotone, \textit{does the full system obey generic convergence properties?}  In other words, suppose that the
flow $\psi_t^0$ of the slow system (set $\varepsilon=0$ in \eqref{eqn:slow_0}):
\begin{align}
\label{eqn:m_0}
\frac{dx}{dt}= f_0(x,m_0(x),0)
\end{align}
has strong monotonicity properties that guarantee almost-global convergence, meaning convergence to equilibria for all initial states except for those states in a set of measure zero (or, in a topological formulation, a nowhere dense set),
but that the complete system is not monotone (so that no such theorem can be applied to it).  Still, one may expect that the almost-convergence result can be lifted to the full system, at least for small $\varepsilon>0$.  An obstruction to this argument is that there is no \emph{a priori} reason for the exceptional set to have a pre-image which has zero measure (or is nowhere dense). Notheless, a positive result along these lines was developed in \cite{wang:JNS07} via the use of  geometric invariant manifold theory to examine the fibration structure and utilize an asymptotic phase property
\cite{Fenichel,Jones,Nipp}.
Figure \ref{fig:M_e} illustrates the idea.
  The ODE
restricted to the invariant manifold $M_\varepsilon $ can be seen as a regular
perturbation of the slow ($\varepsilon $$=$$0$) ODE. In~\cite{Hirsch85}, it has been noted that a $C^1$ regular
perturbation of a flow with  positive derivatives inherits the generic
convergence properties. 
So, solutions in the manifold will generally be well-behaved, and
asymptotic phase means that trajectories close to $M_\varepsilon$ can be approximated by solutions in $M_\varepsilon $,
and hence they also approach the set of equilibria if trajectories on $M_\varepsilon $ do.  
We refer the reader to \cite{wang:JNS07} for a technical discussion. In principle, this approach could be applied to systems such as ours, to conclude almost global convergence, under mild technical conditions on domains of validity for equation.  We omit the details of the application here.
\section{Conclusions}
We have conducted a model reduction analysis for the EMT/MET system. Bounded trajectories of the reduced EMT/MET system generically converge to steady states, assuming sufficiently fast promoter and mRNA kinetics.  Therefore, such a model cannot admit  oscillations nor chaotic behavior. There are many further directions for research, which are being pursued by the authors, including the generalization to the case of miRNA binding to more than two sites on the mRNA molecule. 

 {
\section*{Acknowledgements}
We thank the reviewers of this manuscript for the meticulous reading and the constructive remarks.  This research was supported by NSF grants 1716623 and 1849588.}

\end{document}